\documentclass[conference]{IEEEtran}

\usepackage{cite}
\ifCLASSINFOpdf
  \usepackage[pdftex]{graphicx}
  \graphicspath{{../}{../}}
  \DeclareGraphicsExtensions{.pdf,.jpeg,.png}
\else
   \usepackage[dvips]{graphicx}
  \graphicspath{{../}}
  \DeclareGraphicsExtensions{.eps}
\fi

\begin{document}
%
\title{Adaptive Data Communication Interface: A User-Centric Visual Data Interpretation Framework}

\author{\IEEEauthorblockN{Grazziela P. Figueredo$^{1,2}$, Christian Wagner$^1$, Jonathan M. Garibaldi$^{1,2}$, Uwe Aickelin$^1$, Andr\' e M. S. Barreto$^3$}
\IEEEauthorblockA{1. School of Computer Science, The University of Nottingham, NG8 1BB\\
2. The Advanced Data Analysis Centre, The University of Nottingham, NG8 1BB\\
3. Laborat\' orio Nacional de Computa\c c\~ ao Cient\' ifica, Petr\' opolis\\
gzf, cxw, jmg, uxa@cs.nott.ac.uk, amsb@lncc.br}}

\maketitle

\begin{abstract}
In this position paper, we present ideas about creating a next generation framework towards an adaptive interface for data communication and visualisation systems. Our objective is to develop a system that accepts large data sets as inputs and provides user-centric, meaningful visual information to assist owners to make sense of their data collection. The proposed framework comprises four stages: (i) the knowledge base compilation, where we search and collect existing state-of-the-art visualisation techniques per domain and user preferences; (ii) the development of the learning and inference systems, where we apply artificial intelligence techniques to learn, predict and recommend new graphic interpretations (iii) results evaluation; and (iv) reinforcement and adaptation, where valid outputs are stored in our knowledge base and the system is tuned up to address new demands. These stages, as well as our overall vision, limitations and possible challenges are introduced in this article. We also discuss further extensions of this framework for other knowledge discovery tasks.
\end{abstract}

\IEEEpeerreviewmaketitle

\section{Introduction}

Turning large masses of data into useful information is a multidimensional problem, requiring appropriate data synthesis, analysis, results communication and visualisation. Current techniques generally fall short in delivering effective solutions for these problems, which consequently leads data scientists to struggle with decisions requiring the analysis of big data. New intelligent methods and the alternative employment of existing tools are therefore needed to empower users to explore their data sets. Additionally, different users and knowledge domains have different exploratory needs, and their success in obtaining useful insights from their data depends on how appropriately these requirements are fulfilled. In short, user-centric data knowledge discovery is needed, combining aspects of visualisation, presentation, and interactivity to communicate the analysis results in an appropriate manner. Our interest lies in alleviating the laborious, time-consuming task of creating suitable visual representations to a data content. We therefore introduce an intelligent, extensible, free, open-source framework to assist with data visualisation tasks.

Primarily, data visualization aims at making complex data more accessible and intelligible by accelerating the comprehension of large masses of data summarised in visual abstractions~\cite{Iliinski:2011,Wilkinson:2006}. In many cases, an effective graphic display of the dataset content represents on its own a powerful knowledge discovery tool. Understanding the data content and describing it in some sort of information graphics is a difficult, multi-disciplinary task~\cite{friendly2009milestones}. The information transmitted in a visual must be clear, tailored to the problem and appealing to the viewer. Furthermore, what may sound attractive to a certain profile of viewers or domain area might not be informative to others.

Current available visualisation tools mostly take into consideration solely the dataset metadata to suggest appropriate visualisations~\cite{ReviewVisualisation}, as further described in the next section.  In general, these tools give the user a list of all possible visuals that match the data characteristics. This, however adds a significant burden to the data owners, as they need to understand the charts and choose those that are more appropriate. In addition, users tend to get limited to the visuals they know, i.e. those they have seen or employed in their analysis before. This process is therefore laborious and not necessarily effective.

There are also other categories of visualisation tools offering intelligent support. However, they are mostly proprietary, which makes it difficult for the research community to contribute with their ideas and adapt the tool to their needs~\cite{ReviewVisualisation2}.

From these observations we therefore conclude that there is the need for an adaptable, extensible, open-source suggestion system that takes into consideration (i) the dataset characteristics; (ii) the knowledge domain; (iii) the user profile and preferences; (iv) successful previous visuals matching the problem and domain; (iv) human factors; and (v) possible visual improvements and/or variations to produce what could be an effective way of facilitating the data set knowledge exploration and understanding.

It is our aim therefore to outline a framework encompassing these elements to assist researchers and data owners from several domains to get the best matching graphics to assist interpreting their data set. We are not proposing a tool for visual design; instead, we envision a tool set to store, re-use, infer and evolve existing visual solutions to create effective representation suggestions for new data sets. We hope to implement a mechanism that dynamically finds and executes ``suitable'' representations of the given data set based on a host of bits of information. These representations are then available to the user to pass on, inform their analysis, provide a summary of the data, highlight aspects of the data, etc.

The proposed framework can be divided in four stages. The first phase involves the knowledge base compilation, where a collection of existing state-of-the-art visualisation techniques per domain and user preferences is built. Subsequently, an ensemble of machine learning algorithms is employed to learn and suggest new graphic solutions. These outputs are subsequently evaluated by multiple criteria methods, such as human factors, user ranking, surveys etc. Finally, the evaluation feedback is used to adjust and adapt the interface and feed the original database with additional information. These stages, as well as our overall vision, limitations and possible challenges are further discussed in the remainder of this article.

\section{Related Work}
\label{RelatedWork}

In this section we present a summary of the most relevant features of current visualisation tools from both academia and industry. We have performed an extensive search of existing mechanisms, their functionalities and their application domain. We focused our investigation on understanding how the collaboration between the user and the software occurs. These interactions are of our primary interest, as we want to understand their shortcomings and to propose improvements.

We have therefore divided the existing tools in four main groups, based on how the user interacts with the product: (i) software environments for statistics, computing and graphics; (ii) web-oriented visualisation tools; (iii) interactive tools; and (iv) intelligent interactive tools. Our aim is to discuss the main characteristics of these tools, their advantages, limitations and the existing gaps that culminated in our proposed research. It is not our intention to provide a complete review of existing mechanisms --- instead, we are interested in outlining their common features. For a more extensive review, see~\cite{ReviewVisualisation,ReviewVisualisation2}.

\subsection{Software Environments for Statistics, Computing and Graphics}

This group comprises well-known tools in the scientific community, such as Matlab~\cite{MATLAB:2014}, R~\cite{R}, GNU Octave~\cite{octave:2014}, Python(x,y)~\cite{PythonXY}, etc. They are development environments that allow for data manipulation and provide data visualisations.

These environments are very useful for data analysts, however, they require more effort to learn and use when compared to interactive tools. In addition, scientific programming background is required to produce the data visualisation scripts, which limits their usage to a specific public, such as computer scientists, mathematicians, statisticians, engineers, etc. Furthermore given the profile of the public that uses these tools, the visual charts are mostly tailored to represent research and engineering data.

\subsection{Web-oriented Visualisation Tools}

These tools offer a variety of customisable charts to be displayed in the web, via a web browser. In order to visualise a data set, the user needs to
incorporate the data into scripts, which are generally written in HTML, PhP, JavaScript, etc.

There several available libraries, such as Google Chart~\cite{GoogleCharts}, ZingChart~\cite{ZingChart} and Dygraphs~\cite{Dygraphs}, with a variety of charts for multiple purposes. In addition, there exists various domain-specific tools, such as Instant atlas~\cite{Atltas} (for maps) and Timeline~\cite{Timeline} (for temporal data), etc.

These tools however have the restriction of being web-based, which requires web development expertise and therefore they are not suitable for every user. In addition, having to incorporate the dataset in a web script can be burdensome when dealing with large data.

\subsection{Interactive tools}

Interactive tools enable data owners to develop and share customized data visualizations, from simple charts to advanced graphics. Popular examples are the IBM SPSS Visualisation Designer~\cite{SPSS} and the IBM Many Eyes~\cite{ManyEyes}. The appeal of these tools is that virtually no programming or technical expertise is needed; therefore, almost everyone has the power to create visualisations.

In specific, Many Eyes has several interesting features and operates in line with some of the ideas we are presenting. It creates visualisations from simple data formats, such as text files or spreadsheets. It stores previous data sets so that new users can browse existing visualisations, get ideas and match their problem with existing ones. In addition to their tool, IBM offers expert contributions from a team from the IBM Research’s Center for Advanced Visualisation~\cite{ManyEyes}. This team of experts provides guidance on creating the most effective visualisations and also update user with new trends and advances.

Our intention with this position paper is to add further intelligence to tool sets such as Many Eyes in order to automatise their visualisation suggestions and their expert guidance. In this manner, rather than interacting directly with users, visualisation researchers would communicate primarily with our framework by inputting information and validating results.

\subsection{Intelligent interactive tools}

This group comprises the state-of-the-art of intelligent data visualisation, where additionally to the user interactivity features, system inputs are also offered. Most existing tools are business intelligence-oriented, with focus on a holistic approach for the data analysis rather than just visualisation. A collection of these tools is reviewed in~\cite{ReviewVisualisation2}.

Concerning tools focused only on data visualisation, IBM has recently developed the IBM Rapidly Adaptive Visualization Engine (RAVE)~\cite{ManyEyes}, which is the underlying technology for the most recent version of Many Eyes. RAVE enables the user to describe how his visualizations should look. The technology behind RAVE however is proprietary and therefore we have no information about how the intelligent algorithms used to render the visuals work.

By analysing the existing tools, it is possible to observe that they all have significant shortcomings that impact on data analytics. Some tools require time learning and programming expertise, other tools, such as those where the data needs to be incorporated into a script are only suitable for small data sets; there are environments which are domain-specific; and more sophisticated intelligent tools are not open-source and therefore cannot incorporate changes implemented by the scientific community.

In general, all available tools still fall short in exploring further artificial intelligence methods. Overall, they offer a possibility of visuals and the user chooses one of them to display the data. We envision, instead, intelligent tools that take as input the data set and an array of attributes (metadata, user, domain, etc) and recommends visuals representations of the data. We therefore propose a next generation, innovative framework to fill some of these gaps and to assist users in defining better ways of showing their data graphically. The objective is to create extra intelligent layers in the process of data visualisation to optimise the usefulness of the obtained visuals. Our framework is introduced next.

\section {The Framework}

In this section we introduce our vision and outline the main activities we view as necessary for its accomplishment.  We do not recommend definite solutions for the challenges presented, as some of them are quite complex. Instead, we aim at describing the problem to be addressed and outline a general framework for solving it. Furthermore, one of our objectives is to call other researchers to further explore and develop our proposal.

\begin{figure}[!t]
\centering
\includegraphics[width=3.6in]{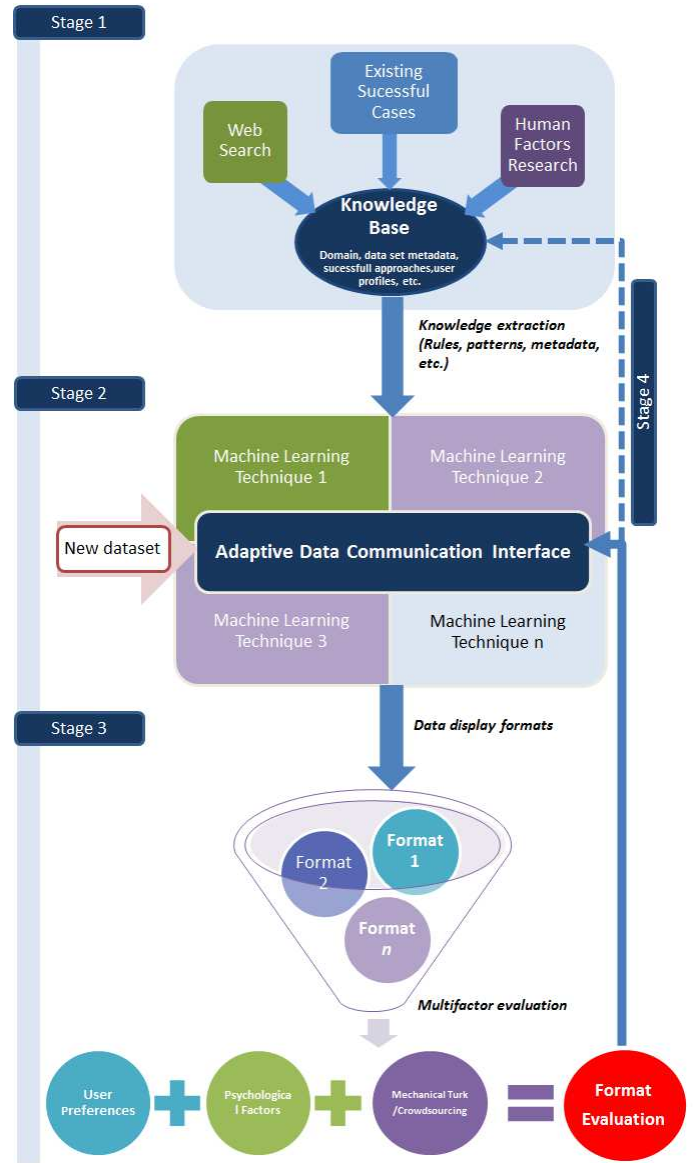}
\caption{The adaptive data communication interface vision}
\label{fig_Framework}
\end{figure}

We envisage a system that takes in a new data set and produces different formats of data visualisation, considering aspects such as the domain area, metadata, user profile and preferences, keywords and previous successful data display cases. The main core of this framework is composed of intelligent methods for knowledge interpretation and inference, to generate and evolve suitable candidate solutions. The outputs produced are evaluated by the target user and other set of criteria; and the results of this evaluation will in turn feed the system with information for future inferences. The scheme of our framework is shown in Figure~\ref{fig_Framework}. There are four main stages to consider, the creation of a knowledge base, the development of the adaptive interface for data communication, the system output evaluation and the feedback. The motivations for the conceptualisation of this idea and each stage of the framework are presented next.

\subsection{Motivations}

The motivations for this framework arose from our work experience. We work with data analytics and we provide analysis for several partners in industry and academia. These partners belong to a variety of areas of expertise, and they have distinct exploratory needs. We are therefore always faced with the challenge of determining the most appropriate graphs to their data sets. In addition, due to budget and time restrictions of our projects, this task needs to be performed quickly and effectively. Our standard procedure when we start characterising the dataset is to produce a number of graphs and assessing their utility, by getting feedback from our clients --- which requires several iterations, each potentially lasting days, delaying the actual analysis.
 
As expected, each domain area and user are drawn to different types of visual interpretations; and we need to learn more about these preferences to ultimately reduce the time spent in the data visualisation task and increase our productivity. We require therefore an intelligent engine that `remembers' successful examples, associates them with tags (domain, user, sub-area, etc.) and recommends visual solutions to new datasets automatically.

\subsection{Framework Stages}

In order to further understand our idea and the motivations behind it, a fictional example is introduced in this section. Let us pretend that we are hired to work on a project involving three professionals, the first one with a Computer Science background, the second one from Biology and the third one from Mathematics. These experts want to investigate the iris dataset. This dataset is available from the UCI Repository of Machine Learning~\cite{umlr:2010} and is a simple, well-known benchmark in the pattern recognition community, which will serve our illustration purposes.  The data set consists of fifty samples from each of three species of iris flowers (Iris {\it setosa}, Iris {\it virginica} and Iris {\it versicolor}). Four features were measured from each sample: the length and the width of the sepals and petals, all in centimetres.

Let us assume the following facts about our researchers (1) the computer scientist is interested in validating a new cluster method; (2) the biologist has never seen this dataset and he is interested in looking at distributions of the species; and (3) the mathematician is interested in creating equations describing the relationship between the petals and sepals of the flowers.

\begin{figure*}[!t]
\centering
\includegraphics[width=7.3in]{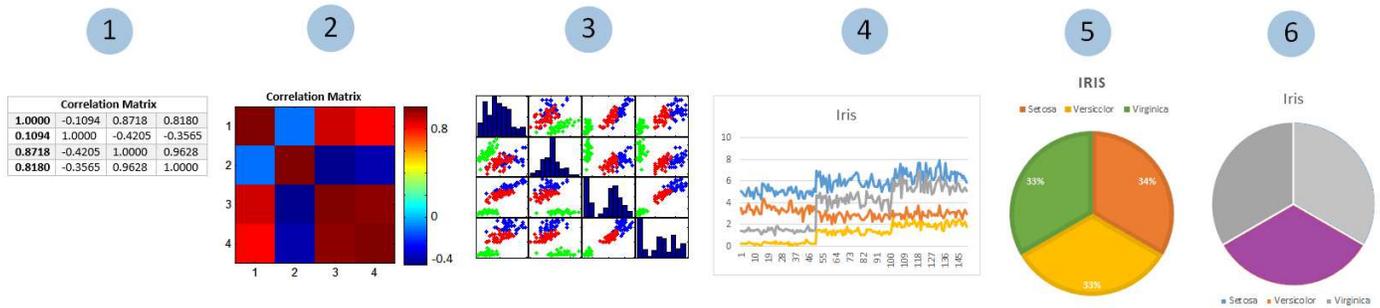}
\caption{Examples of possible visualisations for the Iris dataset}
\label{fig_Iris_Possible_Graphs}
\end{figure*}

Normally, our first step is to assist our partners with some data description and visualisation. We use a set of tools to produce a number of graphs.

Figure~\ref{fig_Iris_Possible_Graphs} shows examples of graphs we produced to interpret the data: (1) a table containing the correlation between the data set attributes; (2) a heat map, which also shows the correlations between the attributes, and therefore represents a different way of displaying the same information as item (1); (3) a scatter plot with histograms, displaying the projections and distributions of the attributes; (4) a line chart, where each line shows the corresponding value (in cm) for each attribute (y axis), for the 150 flowers of the database (x axis); (5) a pie chart with the distribution of flowers per class; and (6) another pie chart showing the same information as graph (5), but with a different design style. All these graphs extract elements of information from the data, however the usefulness of the information transmitted is subjective, depending on the viewers and their questions about the data. 

\begin{figure}[!htpb]
\centering
\includegraphics[width=3.5in]{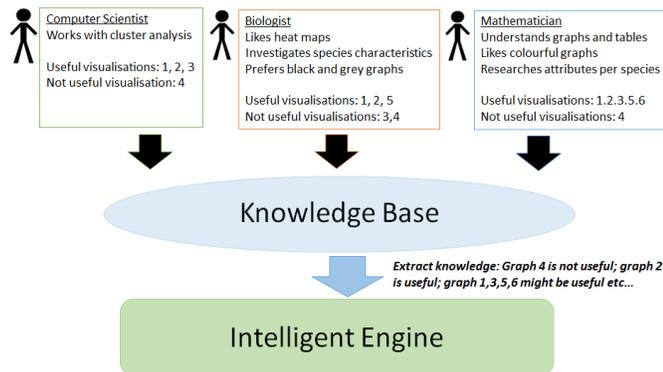}
\caption{Building the knowledge base}
\label{fig_Users}
\end{figure}

In order to determine which are the best visuals, we need feedback from the viewers. Suppose the computer scientist finds graphs (1), (2) and (3) useful for his work and states that visualisation (4) is not useful; the biologist finds (1), (2) and (3) useful and cannot understand figures (3) and (4); and the mathematician understands all graphs, but finds graph (4) not very informative. Once these preferences, associated with the profiles are collected, they need to be stored for future reference. A schematic representation of this process is shown in Figure~\ref{fig_Users}. This part of our example falls into the first stage of our framework (Figure~\ref{fig_Framework}), which is further discussed next.

\subsubsection{Stage 1 - The Knowledge Base Compilation}

As explained in our example, in this stage a ``memory'' of successful cases of data visualisation is built. The purpose is to collect references for the intelligent engine, i.e. our adaptive interface to be able to provide user-centric visual schemes based on previous profiles of similar case studies. This database is supposed to store a maximum possible information about users, domains, types of data, keywords, etc. 

To obtain this knowledge base, an extensive search in the literature, web sites, research papers, online news and open source online data visualisation tools needs to be conducted. In addition, pilot studies and user surveys need to be carried out to determine the human factor elements and requirements that need to be incorporated into the memory mechanism.

This information gathered is subsequently transferred to the intelligent core of our framework in distinct manners, for instance, via direct queries or through extracted knowledge. We foresee a considerable amount of facts being collected, therefore this knowledge base needs to be mined to determine the rules, patterns, tags, parameters, etc. that will serve as inputs to train and tune the adaptive learning system. In addition, once the data is input in the system, it is also likely that it will need some sort of pre-processing and standardisation.

We understand this phase can be quite laborious, and defining `how' and `what' needs to be collected to create our framework memory is a difficult task. In addition, there is the challenge of defining automatic ways and selecting the appropriate technologies to search, collect and store the information needed. Another difficulty that might be encountered is how to define noisy data, i.e. irrelevant information, the misuse of visuals, etc. There is therefore the need to employ different methods of information retrieval~\cite{Manning:2008:IIR:1394399,Turpin:2006:UPV:1148170.1148176,Singhal01moderninformation,Frakes:1992:IRD:129687} and text mining~\cite{CohenH08} in some sort of ensemble to guarantee that, as much as possible, only useful information is stored.

After the knowledge base is created, we subsequently need some sort of intelligent engine that will extract the knowledge collection to feed the next stage.
Data mining needs to be applied to extract rules, patterns, associations, etc to assist building the inference models for new case studies. The collection of inference models constitute the second stage of our framework, as discussed next.




\subsubsection{Stage 2 - The Adaptive Data Communication Interface}

\begin{figure*}[!ht]
\centering
\includegraphics[width=5in]{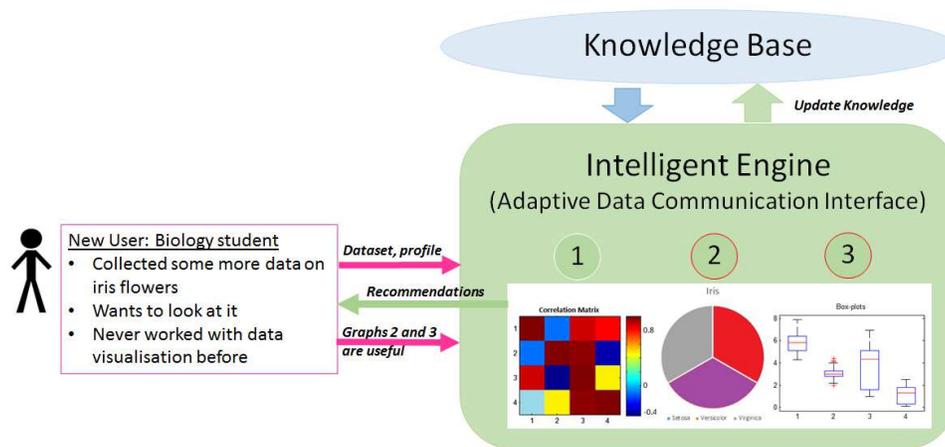}
\caption{The recommendation process}
\label{fig_Recomendation_Process}
\end{figure*}

This is the intelligent nucleus of our framework. The purpose of this interface is to employ the knowledge obtained from past examples to infer and create new graphical interpretations for new problems. According to our ideas, after the knowledge base is created and the intelligent system is trained, the adaptive interface should be able to start providing recommendations. 

Back to our case study, let us suppose a new user has new data and he/she wants to visualise it. The user is a Biology student. He spent his/her internship gathering more instances of the iris flowers and wants to visualise the data collected. He/she inputs his/her profile and dataset to the system, and the intelligent engine needs to work out the best visualisations. Figure~\ref{fig_Recomendation_Process} illustrates a possible manner this engine is supposed to work for this example.

As shown in the figure, the system should output successful graphs that were defined as preferable by similar users. For example, the system could indicate a heat map, a pie chart (based on previous experiences) and recommend a new visual interpretation, i.e. a box-plot (see Figure~\ref{fig_Recomendation_Process}). 

As shown in Figure~\ref{fig_Framework}, we envision this adaptive intelligent interface as a combination of machine learning techniques, such as rule-based approaches~\cite{Piatetsky-Shapiro1991,Agrawal93miningassociation,Hipp00algorithmsfor,Girotra:2013,914856}, case-based reasoning~\cite{Aamodt:1994,Hüllermeier:2007,Schank:1983:DMT:538776,Kolodner1983281}, active learning~\cite{settles2009active,olsson2009literature,Lughofer12a} evolutionary algorithms~\cite{Back:1997:HEC:548530,Back:1996:EAT:229867,Goldberg:1989,Holland:1975:ANA,Michalewicz:1992}, recommendation systems,  and other machine learning methods~\cite{Mitchell:1997,Han:2001}.

ENDS HERE FOR NOW

{\bf Andr\' e, this is the bit I could use some of your brains :)
I want some sort of ensemble, where instead of a majority vote, there would be a majority (or consensus) of what visual would be suggested. (I am not using the word recommendation very often, because I dont want the reader to think this is purely a recommendation system. As discussed, we are creating new visuals, rather than just recommending existing. THis is novel and not part of the traditional recommendation systems (i hope). need to read a bit more about it. Any comments, bullet points very welcome}

As it can be observed... not recomendation system as it generates new unseen cases...

\subsubsection{Stage 3 - Evaluation}

The new user subsequently chooses his favorite graphs, which will in turn feed our knowledge base and adjust our intelligent system.

what the user likes
there could be an online survey also to collect preferences
we could have an input of psychological factors affecting profiles of users. 

this would provide some sort of score or ranking of the existing associations between users, domains etc and the visuals
This would also provide a ranking of those inference models that are not working well. those that need to die, be fixed, those that perform poorly for this task.

this is therefore also a way to evaluate the recommendation/ML techniques used. this will provide an on-line feedback of how our intelligent mechanisms are performing. It will also evaluate the utility of each part of the ensemble.
concerned with testing the framework and a compilation of my multifactor evaluation (user, crowdsourcing, expert evaluation, human and psychological factors) [28-31] results. I intend to perform a detailed appraisal of the environment, outline the insights provided as well as report the types of data and users for which the interface works more appropriately.

\subsubsection{Stage 4 - Adaptation}

\subsection{Discussion}

\section{Conclusion}

Adding intelligence to suggest visualisations is a difficult task. It involves several elements working together to achieve final satisfactory results.

Limitations:

we understand that the domain of data visualisation is and encompassing most visualisation techniques into a single interface seems unachievable. However, the framework can be sucessfuly applied to specific domain areas, within a limited scope. In addition as the system is adaptable, it can migrate to the tecniques that are trendy (useful) for during a certain period. There is also a majority of cases that fall into a definite set of possible visualisations. It is the task of our framework to suggest (assist researchers to define), amongst the toolset, those methods (or set of methods) that combined produce useful visuals.

However, due to the complexity of the problem addressed and the many unforeseen questions that might arise during the implementation of this idea, each stage of the framework on its own represents a large avenue for research and new insights. This framework can possibly be extended to other areas of data analysis.

Our intent, however, is to add an extra interface that could better guide the data owners into the possibilities of exploration of their data.

I understand this is a long term ambitious vision, which might not be entirely accomplished by the end of this fellowship. This is however intended to be ground-breaking research which will lead to a number of long term research activities and further grant proposals.

 EXTENSIONS
 
this is future work
The results of the fellowship grant research will lead to an open-source, extensible framework capable of inputting, processing and outputting results for extremely large datasets.  High performance computing (HPC) based approaches provide a means to effectively manage my interface and thereby improve the interactivity of users of the system leading to a more rapid input-to-result loop. This grant has the aim of developing and adapting my Adaptive Data Communication Interface described below to a HPC environment.


\bibliographystyle{IEEEtran}
\bibliography{Framework}

\end{document}